\documentclass[12pt,letterpaper]{article}
\pdfoutput=1
\usepackage{jheppub}
\usepackage{amsfonts, amsthm}
\usepackage[english]{babel}
\usepackage[utf8]{inputenc}
\usepackage{slashed}
\usepackage{mathrsfs}
\usepackage{amssymb}
\usepackage{xcolor}
\hypersetup{unicode}
\usepackage{lipsum}
\usepackage{breqn}
\usepackage{physics}

\newcommand{\eq}{\begin{equation}}
\newcommand{\feq}{\end{equation}}
\newcommand{\eqn}{\begin{eqnarray}}
\newcommand{\feqn}{\end{eqnarray}}

\newcommand{\mrm}[1]{\mbox{$\mathrm{#1}$}}

\def\SO{{\mathrm{SO}}}
\def\beq{\begin{align}}
\def\eeq{\end{align}}
\newcommand{\bi}{\begin{itemize}}
\newcommand{\ei}{\end{itemize}}
\newcommand{\ben}{\begin{enumerate}}
\newcommand{\een}{\end{enumerate}}
\providecommand{\bea}{\begin{eqnarray}}
\providecommand{\eea}{\end{eqnarray}}
\newcommand{\be}{\begin{equation}}
\newcommand{\ee}{\end{equation}}

\definecolor{amber}{rgb}{1.0, 0.49, 0.0}

\title{New IIB intersecting brane solutions yielding supersymmetric AdS$_3$ vacua}

\author[a]{Juan R. Balaguer}
\author[b,c]{, Giuseppe Dibitetto}
\author[a]{, Jos\'e J. Fern\'andez-Melgarejo}

\affiliation[a]{Departamento de F\'isica, Universidad de Murcia, Campus de Espinardo, E-30100 Murcia, Spain}
\affiliation[b]{Dipartimento di Fisica e Astronomia, Universit\`a di Padova, via Marzolo 8, 35131 Padova, Italy}
\affiliation[c]{INFN, Sezione di Padova, via Marzolo 8, 35131 Padova, Italy}

\emailAdd{juanramon.balaguer@um.es}
\emailAdd{giuseppe.dibitetto@unipd.it}
\emailAdd{melgarejo@um.es}

\abstract{We consider genuine type IIB string theory (supersymmetric) brane intersections that preserve $(1+1)$D Lorentz symmetry. We provide the full supergravity solutions in their analytic form and discuss their physical properties. The Ansatz for the spacetime dependence of the different brane warp factors goes beyond the harmonic superposition principle. By studying the associated near-horizon geometry, we construct interesting classes of AdS$_3$ vacua in type IIB and highlight their relation to the existing classifications in the literature. Finally, we discuss their holographic properties.}

\keywords{}

\begin{document}
\maketitle
\flushbottom

\section{Introduction}
Flux compactifications aim at studying the physical properties of all string theory backgrounds featuring lower-dimensional maximally symmetric geometries. Depending on whether the spacetime curvature is zero, positive, or negative, the corresponding backgrounds represent important playgrounds for string phenomenology. In particular, dS vacua may be used in order to produce toy models for late-time cosmology, while Mkw vacua provide effective models of gravity coupled to (non-)supersymmetric field theories and hence can be very valuable in order to study mechanisms like SUSY breaking. On the other hand, AdS vacua acquired great relevance ever since the very birth of the AdS/CFT correspondence.

Focusing on AdS in particular, an enormous technological progress has been made over the last two decades in the search for vacua preserving some residual supersymmetry. The crucial tool that has been widely employed in this context is the so-called pure spinor formalism \cite{Strominger:1986uh,Gauntlett:2002fz,Grana:2004bg}. It is based on the crucial interplay between background fluxes and geometry whenever SUSY is at least partially preserved in the desired vacuum. This has already allowed us to map out large portions of the string landscape of supersymmetric AdS vacua in various dimensions, revealing interesting structures and providing us with many successful holographic checks, which eventually taught us many things about the AdS/CFT correspondence.

However, an important missing piece in order to go beyond testing holography and put it on more solid grounds is the existence of a brane construction for a given AdS vacuum. The reason for this is that one may appeal to the aforementioned brane construction in order to constructively map the two sides of the correspondence into each other. This is precisely what was successfully carried out in the original work of \cite{Maldacena:1997re} stating the correspondence between type IIB on $\mathrm{AdS}_5\times S^5$ and $\mathcal{N}=4$ SYM$_4$. It may be worth mentioning that, besides a few other examples, the vast majority of known SUSY AdS vacua do not yet admit a known brane picture. The main reason for this is that, the lower the amount of preserved SUSY, the more challenging it gets to find one.

Another advantage of possessing the brane description of a given vacuum is that it gives us the opportunity to study holography even beyond AdS space and hence beyond conformal theories. The first example of this is the so-called DW/QFT correspondence, which was proposed in \cite{Boonstra:1998mp}. This correspondence relates an asymptotically AdS domain wall (DW) to the RG flow of a QFT admitting a conformal fixed point. A crucially different physical situation of this type is that of having an asymptotically AdS \emph{curved} DW (\emph{e.g.} with lower-dimensional AdS slices). Such geometries were proposed in \cite{Clark:2004sb,DHoker:2007zhm} to provide a holographic description of defect CFTs (see also \cite{Gaiotto:2008sd}). This was originally illustrated in the context of Janus solutions describing a conformal interface within $\mathcal{N}=4$ SYM engineered by means of a position-dependent profile for the gauge coupling. On the other hand, a hint towards explaining conformal defects from branes had already been given in \cite{Karch:2000gx,DeWolfe:2001pq}, where it was shown how lower-dimensional AdS interfaces occur in the probe limit when placing defect branes within the background of some \emph{mother} branes admitting a higher-dimensional AdS geometry in the near-horizon limit.
A more explicit relation between these two descriptions of conformal defects was further shown in, \emph{e.g.} \cite{Dibitetto:2017tve,Dibitetto:2017klx,Dibitetto:2018iar,Chen:2019qib,Faedo:2020nol,Dibitetto:2020bsh}, though in the different context of 5d and 6d SCFTs.

A stack of D3 branes has (non-Abelian) $\mathcal{N}=4$ SYM$_4$ as a worldvolume field theory description. This theory enjoys conformal symmetry and maximal rigid supersymmetry. On the supergravity side, an $\mathrm{AdS}_5\times S^5$ geometry emerges when taking the near-horizon limit of the D3 brane metric. By placing additional orthogonal branes on a stack of D3 branes, $(1+3)$D conformal symmetry is broken and one expects a lower-dimensional field theory description to emerge \cite{Harvey:2007ab,Buchbinder:2007ar,Gaiotto:2008sd,Harvey:2008zz}.  In \cite{Choi:2017kxf,Choi:2018fqw}, partially SUSY preserving deformations of $\mathcal{N}=4$ SYM$_4$ with 2-dimensional spacetime dependent profiles for the couplings were studied. It was argued that their stringy origin is the presence of \emph{defect branes} in the background. Consequently, a brane system with D3 branes containing an $\mathrm{AdS}_3$ when taking the near horizon limit is expected as a holographic description of such field theories. 

The aim of this paper is to provide novel brane constructions within a particular setup in type IIB string theory that allow for the supergravity side description of the aforementioned SYM deformations. That is to say, we explore a set of brane intersections (i) containing at least one D3 brane, (ii) with no isometry directions along the orthogonal directions of the D3 brane and (iii) admitting $\mathrm{AdS}_3$ vacua (see \emph{e.g.}\cite{Couzens:2017way,Couzens:2017nnr,Macpherson:2018mif} for wide classifications of well-known classes of these vacua). In particular, some of these vacua preserving $\mathcal{N}=(4,0)$ SUSY are related to (massive) IIA via a T-duality. These latter backgrounds were extensively studied in \cite{Lozano:2019emq,Lozano:2019zvg,Lozano:2019jza,Lozano:2019ywa,Lozano:2020bxo,Faedo:2020lyw}.
We shall rely on the technique developed in \cite{Youm:1999ti} for constructing supergravity solutions describing \emph{semilocalized} brane intersections, \emph{i.e.} where the different brane charge distributions no longer obey the harmonic superposition principle. Such a tool was used in \cite{Boonstra:1998yu} and \cite{Cvetic:2000cj} to engineer AdS vacua with a clear underlying brane interpretation.

The paper is organized as follows. In Sec. \ref{Sec2} we discuss D3 -- D5 -- D7 brane intersections and the form of general solutions in this class. After arguing for the absence of AdS$_3$ near-horizon geometries in this setup, we briefly discuss the interpretation of these solutions as supersymmetric position-dependent profiles for the YM coupling on the D3 branes due to the presence of D5 -- D7 defect branes. In Sec. \ref{Sec3} we move to D3 -- D3 -- D7 intersections. Here we show how an $\mathrm{AdS}_3\times S^3\times\mathbb{T}^2$ geometry with warping over a Riemann surface $\Sigma$ is obtained by taking the near-horizon limit. Further in Sec. \ref{Sec4} we consider a more involved setup, though preserving the same spacetime symmetry as well as the same amount of supersymmetry as the previous cases. The objects intersecting the original stack of D3 branes this time are two differently placed D5s and NS5s, as well as D7 branes. This way, the resulting AdS$_3$ solutions feature a warping over $\mathbb{T}^3\times\Sigma$.
Finally, we refer the reader to appendix \ref{app:iib-string} \& \ref{app:projectors} for some further details concerning our conventions for type IIB supergravity, Killing spinors and SUSY projectors for brane intersections.

\section{D3 -- D5 -- D7 brane systems}
\label{Sec2}

In \cite{Choi:2017kxf}, deformations of $\mathcal{N}=4$ SYM$_4$ with partially SUSY preserving spacetime dependent profiles for the SYM couplings and the mass terms were studied. In particular, the field theory enjoys a rigid $\mathrm{ISO}(1,1)\times\SO(3)\times \SO(3)$ symmetry and the deformation parameters generically depend on two coordinates. It was argued that their stringy origin is the presence of \emph{defect branes} in the background \cite{Choi:2018fqw}.

On the supergravity side, an $\mathrm{AdS}_5\times S^5$ geometry emerges when taking the near-horizon limit of the D3 brane metric. By placing D5 and D7 branes within the aforementioned as depicted in table \ref{Table:D3D5D7}, $(1+3)$D conformal symmetry is broken and one expects a lower-dimensional field theory description to emerge. As explained in appendix \ref{app:projectors}, such a brane intersection turns out to preserve four real supercharges, \emph{i.e.} it is $\frac18$ BPS. Even though no lower-dimensional CFT is expected to emerge in the (new) near-horizon limit, we look for the supergravity description of the field theory construction illustrated in \cite{Choi:2017kxf}.
\begin{table}[http!]
\renewcommand{\arraystretch}{1}
\begin{center}
\scalebox{1}[1]{
\begin{tabular}{c||c c|c c | c c c | c c c }
object & $t$ & $y$ & $x_1$ & $x_2$ & $r$ & $\phi_1$ & $\phi_2$ & $\rho$ & $\theta_{1}$ & $\theta_{2}$      \\
\hline \hline
$\mrm{D}3$ & $\times$ & $\times$ & $\times$ & $\times$ & $\sim$ & $\sim$ & $\sim$ & $-$ & $-$ & $-$   \\
\hline
$\mrm{D}5$& $\times$ & $\times$ & $\times$ & $-$ & $\times$ & $\times$ &  $\times$ & $\sim$ & $\sim$ & $\sim$   \\
$\mrm{D}7$ & $\times$ & $\times$ & $-$ & $-$ & $\times$ & $\times$ & $\times$ & $\times$ & $\times$ & $\times$  \\
\end{tabular}
}
\end{center}
\caption{The $\frac18 $ BPS brane system underlying the intersection of D5 -- D7 branes intersecting D3 branes. The $\sim$ denotes smearing directions.} \label{Table:D3D5D7}
\end{table}
To this end, the 10D supergravity Ansatz we cast is\footnote{We adopt the string frame all throughout the paper. For further information concerning the conventions retained here, we refer to appendix \ref{app:iib-string}.}
\begin{align}
ds_{10}^2=&\, H_{\mathrm{D}3}^{-1/2}H_{\mathrm{D}5}^{-1/2}H_{\mathrm{D}7}^{-1/2}ds^2_{\mathrm{Mkw}_2}
+H_{\mathrm{D}3}^{-1/2}H_{\mathrm{D}7}^{1/2}\left(H_{\mathrm{D}5}^{-1/2}dx_1^2+H_{\mathrm{D}5}^{1/2}dx_2^2\right)
\nonumber\\
+& \, H_{\mathrm{D}3}^{1/2}H_{\mathrm{D}5}^{-1/2}H_{\mathrm{D}7}^{-1/2}\left(dr^2+r^2 ds^2_{S^2}\right)
+H_{\mathrm{D}3}^{1/2}H_{\mathrm{D}5}^{1/2}H_{\mathrm{D}7}^{-1/2}\left(d\rho^2+\rho^2ds^2_{\tilde{S}^2}\right)
\ , \\[1mm]
C_{(4)} =&\ H_{\mathrm{D}7} H_{\mathrm{D}3}^{-1}\,\mathrm{vol}_{\mathrm{Mkw}_2}\wedge dx_1 \wedge dx_2|_{\mathrm{SD}}\ ,\\[1mm]
C_{(6)} =&\ H_{\mathrm{D}5}^{-1}\,\mathrm{vol}_{\mathrm{Mkw}_2}\wedge dx_1 \wedge \mathrm{vol}_{\mathbb{R}^3}\ , \\[1mm]
C_{(8)} =&\  H_{\mathrm{D}3}H_{\mathrm{D}7}^{-1}\,\mathrm{vol}_{\mathrm{Mkw}_2}\wedge \mathrm{vol}_{\mathbb{R}^3} \wedge \mathrm{vol}_{\tilde{\mathbb{R}}^3}\ ,
\\[1mm]
e^{\Phi}=&\ H_{\mathrm{D}5}^{-1/2}H_{\mathrm{D}7}^{-1}\ ,
\end{align}
where $ds^2_{\mathrm{Mkw}_2}\equiv(-dt^2+dy^2)$, $ds^2_{S^2}\equiv(d\phi_1^2+\sin^2\phi_1 d\phi_2^2)$, $ds^2_{\tilde{S}^2}\equiv(d\theta_1^2+\sin^2\theta_1 d\theta_2^2)$, and $(\cdots)|_{\mathrm{SD}}$ denotes projection onto the self-dual field.
The warp factors appearing in the above Ansatz are respectively assumed to have the following spacetime dependence: $H_{\mathrm{D}3}=H_{\mathrm{D}3}(\rho)$, $H_{\mathrm{D}5}=H_{\mathrm{D}5}(x_2)$ and $H_{\mathrm{D}7}=H_{\mathrm{D}7}(x_1,x_2)$.

The complete set of BI \eqref{BIH3} -- \eqref{BIG9} and equations of motion \eqref{EOMPhi} -- \eqref{EinsteinEQ} is satisfied upon imposing the following differential equations
\begin{align}
\partial_{x_2}^2H_{\mathrm{D}5} \overset{!}{=} & \ 0\ , \label{HD5''=0}
\\
\Delta_{\tilde{\mathbb{R}}^3}H_{\mathrm{D}3} \ \equiv \ \partial_{\rho}^2H_{\mathrm{D}3}+\frac2\rho \partial_{\rho}H_{\mathrm{D}3} \overset{!}{=} & \ 0\ , \label{DeltaHD3=0}
\\
H_{\mathrm{D}5} \,\partial_{x_1}^2 H_{\mathrm{D}7}+\partial_{x_2}^2H_{\mathrm{D}7}\overset{!}{=} & \ 0\ . \label{DeltaHD7=0}
\end{align}
It may be worth mentioning that, while the first two differential conditions impose harmonicity and can be integrated in fully generality, the third one admits \emph{non-harmonic} solutions for $H_{\mathrm{D}7}$, and in particular its explicit form crucially depends on the choice of $H_{\mathrm{D}5}$.

Equation \eqref{HD5''=0} is solved by
\begin{equation}
H_{\mathrm{D}5} \ = \ h_{\mathrm{D}5} \, + \, Q_{\mathrm{D}5} x_2 \ ,
\end{equation}
where $h_{\mathrm{D}5}$ \& $Q_{\mathrm{D}5}$ are some real integration constants, while \eqref{DeltaHD3=0} yields
\begin{equation}
H_{\mathrm{D}3} \ = \ h_{\mathrm{D}3} \, + \, \frac{Q_{\mathrm{D}3}}{\rho} \ ,
\end{equation}
$h_{\mathrm{D}3}$ \& $Q_{\mathrm{D}3}$ being yet new real integration constants. The last condition to determine where $H_{\mathrm{D}7}$ is generically only solvable term by term in its Laurent expansion. However, there are special cases that allow for a resummation of the resulting solution. For example, when the D5s are nearly cored (\emph{i.e.} $h_{\mathrm{D}5}=0$), \eqref{DeltaHD7=0} is solved by
\begin{align}
H_{\mathrm{D}7} \ = \ \left\{\begin{array}{lc}h_{\mathrm{D}7}+Q_{\mathrm{D}7}\left(x_1^2-\dfrac{Q_5}{3}x_2^3\right) & ,\\ \textrm{or} & , \\ h_{\mathrm{D}7}+Q_{\mathrm{D}7}\left(x_1^2+\dfrac{4Q_5}{9} x_2^3\right)^{-1/6} & , \end{array}\right.
\end{align}
in terms of two extra integration constants $h_{\mathrm{D}7}$ \& $Q_{\mathrm{D}7}$. When $h_{\mathrm{D}5}\neq 0$, simple solutions are of course found by solving $\partial_{x_1}^2 H_{\mathrm{D}7}=0$ and $\partial_{x_2}^2 H_{\mathrm{D}7}=0$ separately. Otherwise, more non-trivial possibilities are
\begin{align}
H_{\mathrm{D}7} \ = \ H^{(0)}_{\mathrm{D}7} \, + \, \frac{x_1^2}{2} \, - \, H_{\mathrm{D}5}^{(-2)} \ ,
\end{align}
where the function $H^{(0)}_{\mathrm{D}7}$ satisfies $\partial_{x_1}^2H^{(0)}_{\mathrm{D}7}\,=\,\partial_{x_2}^2H^{(0)}_{\mathrm{D}7}\,=\,0$, and $H_{\mathrm{D}5}^{(-2)}$ denotes the second primitive of $H_{\mathrm{D}5}$, \emph{i.e.} satisfying $\partial_{x_2}^2H^{(-2)}_{\mathrm{D}5}\,=\,H_{\mathrm{D}5}$.

The class of solutions presented here does not contain any obvious AdS$_3$ solutions to be obtained by means of an appropriate near-horizon procedure. Nevertheless, they have a possible interesting physical interpretation as the gravity duals of the position dependent configurations of the coupling within $\mathcal{N}=4$ SYM, which were studied in \cite{Choi:2018fqw}.

\section{D3 -- D3$'$ -- D7 brane systems}
\label{Sec3}
Let us now move to a different brane configuration in type IIB. Motivated by the SYM deformation with $\mathrm{ISO}(1,1)\times\SO(2)\times\SO(4)$ symmetry and 2D spacetime dependent couplings obtained in \cite{Choi:2017kxf}, we propose the following $\frac18$-BPS brane configuration.

Within the same D3 brane background, we place this time orthogonal D3 branes (denoted by D$3'$) as well as D7 branes in such a way as to leave $(1+1)$D Lorentz symmetry intact. This procedure, which is illustrated in detail in table \ref{Table:D3D3D7}, turns out to preserve four real supercharges\footnote{We refer again to appendix \ref{app:projectors} for a more detailed explanation concerning the SUSY projectors.}.  
\begin{table}[http!]
\renewcommand{\arraystretch}{1}
\begin{center}
\scalebox{1}[1]{
\begin{tabular}{c||c c|c c | c c | c c c c }
object & $t$ & $y$ & $x_1$ & $x_2$ & $r$ & $\phi$ &  $\rho$ & $\theta_{1}$ & $\theta_{2}$   & $\theta_{3}$   \\
\hline \hline
$\mrm{D}3$ & $\times$ & $\times$ & $\times$ & $\times$ & $-$ & $-$ & $-$ & $-$ & $-$ & $-$   \\
\hline
$\mrm{D}3'$& $\times$ & $\times$ & $\sim$ & $\sim$ & $\times$ & $\times$ &  $-$ & $-$ & $-$ & $-$   \\
$\mrm{D}7$ & $\times$ & $\times$ & $-$ & $-$ & $\times$ & $\times$ & $\times$ & $\times$ & $\times$ & $\times$  \\
\end{tabular}
}
\end{center}
\caption{The $\frac18 $ BPS brane system underlying the intersection of D3$'$ -- D7 branes intersecting D3 branes. The $\sim$ denotes smearing directions.} \label{Table:D3D3D7}
\end{table}
\begin{align}
ds_{10}^2=&\, H_{\mathrm{D}3}^{-1/2}H_{\mathrm{D}3'}^{-1/2}H_{\mathrm{D}7}^{-1/2}ds^2_{\mathrm{Mkw}_2}
+H_{\mathrm{D}3}^{-1/2}H_{\mathrm{D}3'}^{1/2}H_{\mathrm{D}7}^{1/2}\left(dx_1^2+dx_2^2\right)
\nonumber\\
+& \, H_{\mathrm{D}3}^{1/2}H_{\mathrm{D}3'}^{-1/2}H_{\mathrm{D}7}^{-1/2}\left(dr^2+r^2 d\phi^2\right)
+H_{\mathrm{D}3}^{1/2}H_{\mathrm{D}3'}^{1/2}H_{\mathrm{D}7}^{-1/2}\left(d\rho^2+\rho^2ds^2_{{S}^3}\right)
\ , \label{D3D3D7_metric} \\[1mm]
C_{(4)} =&\ \left(H_{\mathrm{D}7} H_{\mathrm{D}3}^{-1}\,\mathrm{vol}_{\mathrm{Mkw}_2}\wedge dx_1 \wedge dx_2+H_{\mathrm{D}3'}^{-1}\,\mathrm{vol}_{\mathrm{Mkw}_2}\wedge \mathrm{vol}_{\mathbb{R}^2}\right)|_{\mathrm{SD}}\ ,\\[1mm]
C_{(6)} =&\ 0 \ , \\[1mm]
C_{(8)} =&\  H_{\mathrm{D}3}H_{\mathrm{D}7}^{-1}\,\mathrm{vol}_{\mathrm{Mkw}_2}\wedge \mathrm{vol}_{\mathbb{R}^2} \wedge \mathrm{vol}_{{\mathbb{R}}^4}\ ,
\\[1mm]
e^{\Phi}=&\ H_{\mathrm{D}7}^{-1}\ , \label{D3D3D7_dilaton}
\end{align}
where $ds^2_{\mathrm{Mkw}_2}\equiv(-dt^2+dy^2)$,  $ds^2_{{S}^3}\equiv(d\theta_1^2+\sin^2\theta_1 d\theta_2^2+\sin^2\theta_1 \sin^2\theta_2 d\theta_3^2)$, and $(\cdots)|_{\mathrm{SD}}$ denotes projection onto the self-dual field.
The warp factors appearing in the above Ansatz are respectively assumed to have the following spacetime dependence: $H_{\mathrm{D}3}=H_{\mathrm{D}3}(r,\rho)$, $H_{\mathrm{D}3'}=H_{\mathrm{D}3'}(\rho)$ and $H_{\mathrm{D}7}=H_{\mathrm{D}7}(x_1,x_2)$.

The complete set of BI \eqref{BIH3} -- \eqref{BIG9} and equations of motion \eqref{EOMPhi} -- \eqref{EinsteinEQ} is satisfied upon imposing the following differential equations
\begin{align}
\Delta_{{\mathbb{R}}^4}H_{\mathrm{D}3} +H_{\mathrm{D}3'}\Delta_{{\mathbb{R}}^2}H_{\mathrm{D}3} \ \equiv \ 
\rho^{-3}\partial_{\rho}\left(\rho^3\partial_{\rho}H_{\mathrm{D}3}\right)+H_{\mathrm{D}3'}r^{-1}\partial_{r}\left(r\partial_{r}H_{\mathrm{D}3}\right)\overset{!}{=} & \ 0\ , \label{DeltaHD3New=0}
\\[1mm]
\Delta_{{\mathbb{R}}^4}H_{\mathrm{D}3'} \ \equiv \ \rho^{-3}\partial_{\rho}\left(\rho^3\partial_{\rho}H_{\mathrm{D}3'}\right) \overset{!}{=} & \ 0\ , \label{DeltaHD3'=0}
\\[1mm]
\Delta_{\Sigma}H_{\mathrm{D}7} \ \equiv \ \partial_{x_1}^2 H_{\mathrm{D}7}+\partial_{x_2}^2H_{\mathrm{D}7}\overset{!}{=} & \ 0\ . \label{DeltaHD7New=0}
\end{align}
In this case the above system can be \emph{e.g.} solved by\footnote{Note that, depending on possible BCs chosen in the $(x_1,x_2)$ plane, $H_{\mathrm{D}7}$ can be an arbitrary harmonic function. The one we write here is the only rotationally invariant form for this function, but there are infinite different solutions.}
\begin{align}
H_{\mathrm{D}3} = \ & 1+Q_{\mathrm{D}3}\left(\frac{1}{\rho^2}+\kappa_1\log r+\kappa_2\log \rho+\kappa_3\left(\rho^2-2 r ^2\right)\right)\ ,
\\
H_{\mathrm{D}3'} = \ & 1+\frac{Q_{\mathrm{D}3'}}{\rho^2}\ ,
\\
H_{\mathrm{D}7} = \ & 1+Q_{\mathrm{D}7}\log(x_1^2+x_2^2)\ ,
\end{align}
where $Q_{\mathrm{D}3}$, $Q_{\mathrm{D}3'}$, $Q_{\mathrm{D}7}$, $\kappa_i$ are real integration constants satisfying
\begin{equation}
\kappa_2 \ \overset{!}{=} \ 4 Q_{\mathrm{D}3'} \kappa_3 \ .
\end{equation}
\subsection*{Near-Horizon limit and AdS$_3$}
As $\rho\rightarrow 0$, one has the following asymptotic behavior for the warp factors
\begin{align}
\begin{array}{lclclc}
H_{\mathrm{D}3} \ \sim \ \dfrac{Q_{\mathrm{D}3}}{\rho^2} & , & \textrm{and} & & H_{\mathrm{D}3'} \ \sim \ \dfrac{Q_{\mathrm{D}3'}}{\rho^2} & ,
\end{array}
\end{align}
while $H_{\mathrm{D}7}$ stays finite since it is a $\rho$-independent harmonic function of $(x_1,x_2)$. In this limit, the solution in  \eqref{D3D3D7_metric} --  \eqref{D3D3D7_dilaton}, after rescaling the Minkowski worldvolume
\footnote{We take $(t,y)\to \frac{1}{\ell}(t,y)$.}, takes the form
\begin{align}
ds_{10}^2=&\, \ell^2 H^{-1/2}\left(ds^2_{\mathrm{AdS}_3}
+Q_{\mathrm{D}3}^{-1} H ds_{\Sigma}^2
+ Q_{\mathrm{D}3'}^{-1} ds_{\mathbb{T}^2}^2
+ds^2_{{S}^3}\right)
\ , \label{AdS31_metric} \\[1mm]
C_{(4)} =&\ \ell^2\rho^2\,\mathrm{vol}_{\mathrm{Mkw}_2}\wedge \left({Q_{\mathrm{D}3}}^{-1}H\mathrm{vol}_{\Sigma}+Q_{\mathrm{D}3'}^{-1}\mathrm{vol}_{\mathbb{T}^2}\right)|_{\mathrm{SD}}\ ,\\[1mm]
C_{(6)} =&\ 0 \ , \\[1mm]
C_{(8)} =&\ \ell^4 Q_{\mathrm{D}3}H^{-1}\,\mathrm{vol}_{\mathrm{AdS}_3}\wedge \mathrm{vol}_{\mathbb{T}^2} \wedge \mathrm{vol}_{S^3}\ ,
\\[1mm]
e^{\Phi}=&\ H^{-1}\ , \label{AdS31_dilaton}
\end{align}
where $ds^2_{\mathrm{AdS}_3}\equiv \left(\frac{\rho^2}{\ell^2}ds^2_{\mathrm{Mkw}_2}+\frac{d\rho^2}{\rho^2}\right)$ is the metric of unit AdS$_3$, with $\mathrm{vol}_{\mathrm{AdS}_3}\equiv \ell^{-2} \rho\, dt\wedge dy\wedge d\rho$ the volume element, $ds_{\Sigma}^2\equiv\left(dx_1^2+dx_2^2\right)$ and  $ ds_{\mathbb{T}^2}^2\equiv\left(dr^2+r^2 d\phi^2\right)$, while the constant length $\ell$ is defined as $\ell\equiv (Q_{\mathrm{D}3}Q_{\mathrm{D}3'})^{1/4}$, and finally $H$ is an arbitrary harmonic function on $\Sigma$.

The above solution represents $\mathrm{AdS}_3\times S^3\times\mathbb{T}^2$ warped over a Riemann surface $\Sigma$. As usual in these cases, besides an enhancement of spacetime symmetry, such a near-horizon geometry also possesses \emph{enhanced supersymmetry}, from $4$ to $8$ real supercharges. 

\subsection*{Discussion}

When turning off a D3 brane charge, namely either $H_{\mathrm{D}3}$ or $H_{\mathrm{D}3'}$ are constant functions, a global symmetry enhancement to an $S^5$ takes place and the resulting D3--D7 brane intersection exhibits the near horizon geometry $\mathrm{AdS}_3\times S^5\times \mathbb{T}^2$. Precisely, a particular choice of the deformation parameters of the $\SO(2)\times\SO(4)$ field theory obtained in \cite{Choi:2017kxf} allows for such symmetry enhancement, in such a way that the resulting action exhibits a rigid $\mathrm{ISO}(1,1)\times\SO(6)$ invariance  \cite{Harvey:2007ab,Buchbinder:2007ar,Harvey:2008zz}.

Similarly, when turning off the D7 brane charge, the $\SO(2)\times\SO(4)$ remains and there is a SUSY enhancement to 16 supercharges \cite{DHoker:2008wvd}. These solutions have been studied in \cite{Gomis:2007fi,Drukker:2008wr}.

Despite a D3 -- D7 -- D7$'$ construction with the same rigid symmetry has been explored, we have not found a supergravity solution without smearing along the $(x_1,x_2)$ directions.

In light of the classification for probe branes in $\mathrm{AdS}_5\times S^5$ done in \cite{DHoker:2008wvd}, it would be interesting to study a stack of $N$ D3 branes within a D3$'$ -- D7 background in the probe approximation.

\section{D3 -- D5 -- NS5 -- D5$'$ -- NS5$'$ -- D7 brane systems}
\label{Sec4}
Let us now return to the first brane system that we considered. We argued that it did not admit any special near-horizon geometry featuring lower-dimensional AdS factors. However, it was already observed in \cite{Faedo:2020lyw} that more general intersections of this type do yield AdS$_3$ vacua in Type IIB supergravity with $\mathcal{N}=2$ supersymmetry in 3D. The explicit brane setup considered there is given in table \ref{Table:D3D5NS5D5NS5D7} and it is originally obtained by performing a single T-duality on the (massive) Type IIA setup given by D2 -- D6 -- NS5 -- D4 -- D8 -- KK5. In \cite{Faedo:2020lyw} both the supergravity brane solution and the dual 2d quivers are discussed.
\begin{table}[http!]
\renewcommand{\arraystretch}{1}
\begin{center}
\scalebox{1}[1]{
\begin{tabular}{c||c c|c c | c c c | c c c }
object & $t$ & $y$ & $x_1$ & $x_2$ & $r$ & $\phi_1$ & $\phi_2$ & $\rho$ & $\theta_{1}$ & $\theta_{2}$      \\
\hline \hline
$\mrm{D}3$ & $\times$ & $\times$ & $\times$ & $\times$ & $-$ & $-$ & $-$ & $-$ & $-$ & $-$   \\
\hline
$\mrm{D}5$& $\times$ & $\times$ & $\sim$ & $\times$ & $-$ & $-$ &  $-$ & $\times$ & $\times$ & $\times$   \\
$\mrm{NS}5$& $\times$ & $\times$ & $\times$ & $\sim$ & $-$ & $-$ & $-$ &  $\times$ & $\times$ &  $\times$  \\
$\mrm{D}5'$& $\times$ & $\times$ & $\times$ & $\sim$ & $\times$ & $\times$ &  $\times$ & $-$ & $-$ & $-$   \\
$\mrm{NS}5'$& $\times$ & $\times$ & $\sim$ & $\times$ & $\times$ & $\times$ &  $\times$ & $-$ & $-$ & $-$   \\
$\mrm{D}7$ & $\times$ & $\times$ & $-$ & $-$ & $\times$ & $\times$ & $\times$ & $\times$ & $\times$ & $\times$  \\
\end{tabular}
}
\end{center}
\caption{The $\frac18 $ BPS brane system underlying the intersection of D5 -- NS5 -- D5$'$ -- NS5$'$ -- D7 branes intersecting D3 branes. The $\sim$ denotes smearing directions.} \label{Table:D3D5NS5D5NS5D7}
\end{table}

In this paper we want to consider this very same brane setup, but, as opposed to \cite{Faedo:2020lyw}, we do not want to keep any link with the (massive) IIA description. As a result, we will obtain a generalization of the brane solution described there, where no isometric circle is left in order to perform the T-duality back to Type IIA. Our solution will be \emph{genuinely} Type IIB. Our Ansatz on the IIB fields reads
\begin{align}
ds_{10}^2=&\, H_{\mathrm{D}3}^{-1/2}H_{\mathrm{D}5}^{-1/2}H_{\mathrm{D}5'}^{-1/2}H_{\mathrm{D}7}^{-1/2}ds^2_{\mathrm{Mkw}_2}
+H_{\mathrm{D}3}^{-1/2}H_{\mathrm{D}7}^{1/2}\left(\frac{H_{\mathrm{D}5'}^{1/2}}{H_{\mathrm{D}5}^{1/2}}H_{\mathrm{NS}5}dx_1^2+\frac{H_{\mathrm{D}5}^{1/2}}{H_{\mathrm{D}5'}^{1/2}}H_{\mathrm{NS}5'}dx_2^2\right)
\nonumber\\
+& \, H_{\mathrm{D}3}^{1/2}H_{\mathrm{D}7}^{-1/2}\left(\frac{H_{\mathrm{D}5}^{1/2}}{H_{\mathrm{D}5'}^{1/2}}H_{\mathrm{NS}5}\left(dr^2+r^2 ds^2_{S^2}\right)
+\frac{H_{\mathrm{D}5'}^{1/2}}{H_{\mathrm{D}5}^{1/2}}H_{\mathrm{NS}5'}\left(d\rho^2+\rho^2ds^2_{\tilde{S}^2}\right)\right)
\ , \label{D3D5NS5D5NS5D7_metric}\\[1mm]
C_{(4)} =&\ H_{\mathrm{D}7} H_{\mathrm{D}3}^{-1}\,\mathrm{vol}_{\mathrm{Mkw}_2}\wedge dx_1 \wedge dx_2|_{\mathrm{SD}}\ ,\\[1mm]
C_{(6)} =&\ \frac{H_{\mathrm{D}5'}H_{\mathrm{NS}5'}}{H_{\mathrm{D}5}}\,\mathrm{vol}_{\mathrm{Mkw}_2}\wedge dx_1 \wedge \mathrm{vol}_{\tilde{\mathbb{R}}^3}-\frac{H_{\mathrm{D}5}H_{\mathrm{NS}5}}{H_{\mathrm{D}5'}}\,\mathrm{vol}_{\mathrm{Mkw}_2}\wedge dx_2 \wedge \mathrm{vol}_{\mathbb{R}^3}\ , \\[1mm]
C_{(8)} =&\  H_{\mathrm{D}3}H_{\mathrm{NS}5}H_{\mathrm{NS}5'}H_{\mathrm{D}7}^{-1}\,\mathrm{vol}_{\mathrm{Mkw}_2}\wedge \mathrm{vol}_{\mathbb{R}^3} \wedge \mathrm{vol}_{\tilde{\mathbb{R}}^3}\ ,
\\[1mm]
B_{(6)} =&\ H_{\mathrm{D}7}\left(\frac{H_{\mathrm{D}5}H_{\mathrm{NS}5}}{H_{\mathrm{NS}5'}}\,\mathrm{vol}_{\mathrm{Mkw}_2}\wedge dx_1 \wedge \mathrm{vol}_{\mathbb{R}^3}+\frac{H_{\mathrm{D}5'}H_{\mathrm{NS}5'}}{H_{\mathrm{NS}5}}\,\mathrm{vol}_{\mathrm{Mkw}_2}\wedge dx_2 \wedge \mathrm{vol}_{\tilde{\mathbb{R}}^3}\right)\ ,
\\[1mm]
e^{\Phi}=&\ H_{\mathrm{D}5}^{-1/2}H_{\mathrm{D}5'}^{-1/2}H_{\mathrm{NS}5}^{1/2}H_{\mathrm{NS}5'}^{1/2}H_{\mathrm{D}7}^{-1}\ ,
\label{D3D5NS5D5NS5D7_dilaton}
\end{align}
where $ds^2_{\mathrm{Mkw}_2}\equiv(-dt^2+dy^2)$, $ds^2_{S^2}\equiv(d\phi_1^2+\sin^2\phi_1 d\phi_2^2)$, $ds^2_{\tilde{S}^2}\equiv(d\theta_1^2+\sin^2\theta_1 d\theta_2^2)$, and $(\cdots)|_{\mathrm{SD}}$ denotes projection onto the self-dual field.
The warp factors appearing in the above Ansatz are respectively assumed to have the following spacetime dependence: $H_{\mathrm{D}3}=H_{\mathrm{D}3}(r,\rho)$, $H_{\mathrm{D}5}=H_{\mathrm{D}5}(r)$, $H_{\mathrm{NS}5}=H_{\mathrm{NS}5}(r)$, $H_{\mathrm{D}5'}=H_{\mathrm{D}5'}(\rho)$, $H_{\mathrm{NS}5'}=H_{\mathrm{NS}5'}(\rho)$ and $H_{\mathrm{D}7}=H_{\mathrm{D}7}(x_1,x_2)$.

In order for the complete set of BI \eqref{BIH3} -- \eqref{BIG9} and equations of motion \eqref{EOMPhi} -- \eqref{EinsteinEQ} to be satisfied, the following identification turns out to be necessary\footnote{It is worth noticing that these conditions represent the most general way of solving for the dynamics of the system if one aims at retaining a completely general $(x_1,x_2)$ dependence of $H_{\mathrm{D}7}$.}
\begin{equation}\label{identification}
H_{\mathrm{NS}5} \ = \ H_{\mathrm{D}5} \ , \qquad \textrm{and} \qquad \ H_{\mathrm{NS}5'} \ = \ H_{\mathrm{D}5'} \ .
\end{equation}
After taking \eqref{identification} into account, one is left with the following set of differential equations
\begin{align}
\Delta_{\tilde{\mathbb{R}}^3}H_{\mathrm{D}3} +\frac{H_{\mathrm{D}5'}^2}{H_{\mathrm{D}5}^2}\Delta_{{\mathbb{R}}^3}H_{\mathrm{D}3} \ \equiv \ 
\rho^{-2}\partial_{\rho}\left(\rho^2\partial_{\rho}H_{\mathrm{D}3}\right)+\frac{H_{\mathrm{D}5'}^2}{H_{\mathrm{D}5}^2}r^{-2}\partial_{r}\left(r^2\partial_{r}H_{\mathrm{D}3}\right)\overset{!}{=} & \ 0\ , \label{DeltaHD3Complete=0}
\\[1mm]
\Delta_{{\mathbb{R}}^3}H_{\mathrm{D}5} \ \equiv \ r^{-2}\partial_{r}\left(r^2\partial_{r}H_{\mathrm{D}5}\right) \overset{!}{=} & \ 0\ , \label{HD5Harm}
\\[1mm]
\Delta_{{\tilde{\mathbb{R}}}^3}H_{\mathrm{D}5'} \ \equiv \ \rho^{-2}\partial_{\rho}\left(\rho^2\partial_{\rho}H_{\mathrm{D}5'}\right) \overset{!}{=} & \ 0\ , 
\label{HD5'Harm}\\[1mm]
\Delta_{\Sigma}H_{\mathrm{D}7} \ \equiv \ \partial_{x_1}^2 H_{\mathrm{D}7}+\partial_{x_2}^2H_{\mathrm{D}7}\overset{!}{=} & \ 0\ . 
\end{align}
Equations \eqref{DeltaHD3Complete=0}, \eqref{HD5Harm} and \eqref{HD5'Harm} are respectively  solved by
\begin{align}
H_{\mathrm{D}3}  =\ & 1+Q_{\mathrm{D}3}\left(\frac1r+\frac\kappa\rho\right)\ ,\\
H_{\mathrm{D}5}  =\ & H_{\mathrm{NS}5}  =\ 1+\frac{Q_{5}}{r}\ ,\\
H_{\mathrm{D}5'}  =\ & H_{\mathrm{NS}5'}  =\ 1+\frac{Q'_{5}}{\rho}\ ,
\end{align}
while we leave $H_{\mathrm{D}7}$ as an unspecified harmonic function on $\Sigma$.
\subsection*{Near-Horizon limit and AdS$_3$}
As $r\rightarrow 0$ (or identically $\rho\rightarrow 0$, since the whole solution is completely symmetric w.r.t. swapping $r\leftrightarrow\rho$), one has the following asymptotic behavior for the warp factors
\begin{align}
\begin{array}{lclclc}
H_{\mathrm{D}3} \ \sim \ \dfrac{Q_{\mathrm{D}3}}{r} & , & \textrm{and} & & H_{\mathrm{D}5} \,=\,  H_{\mathrm{NS}5} \ \sim \ \dfrac{Q_{5}}{r} & ,
\end{array}
\end{align}
while $H_{\mathrm{D}5'}$, $H_{\mathrm{NS}5'}$ and $H_{\mathrm{D}7}$ stay finite since they are $r$-independent harmonic functions on $\tilde{\mathbb{R}}^3$ and $\Sigma$, respectively. In this limit, the solution in  \eqref{D3D5NS5D5NS5D7_metric} --  \eqref{D3D5NS5D5NS5D7_dilaton}, after rescaling the Minkowski worldvolume
\footnote{We take $(t,y)\to \frac{1}{2\ell}(t,y)$.}, takes the form
\begin{align}
ds_{10}^2=&\, \ell^2 H^{-1/2}K^{-1/2}\left(4 ds^2_{\mathrm{AdS}_3}
+ \frac{H K}{Q_{\mathrm{D}3}Q_5} ds_{\Sigma}^2
+ \frac{K^2}{Q_{5}^{2}} ds_{\mathbb{T}^3}^2
+ds^2_{{S}^2}\right)
\ , \label{AdS32_metric} \\[1mm]
C_{(4)} =&\ 4\ell^2{Q_{\mathrm{D}3}}^{-1} H r\,\mathrm{vol}_{\mathrm{Mkw}_2}\wedge\mathrm{vol}_{\Sigma}|_{\mathrm{SD}}\ ,\\[1mm]
C_{(6)} =&\ 4\ell^2 {Q_5}^{-1} K^2 r \, \mathrm{vol}_{\mathrm{Mkw}_2}\wedge dx_1\wedge\mathrm{vol}_{\mathbb{T}^3}+8\ell^4{Q_5} K^{-1}\mathrm{vol}_{\mathrm{AdS}_3}\wedge dx_2 \wedge \mathrm{vol}_{S^2}\ , \\[1mm]
C_{(8)} =&\ 8\ell^4{Q_{\mathrm{D}3}} H^{-1}K \,\mathrm{vol}_{\mathrm{AdS}_3}\wedge \mathrm{vol}_{S^2}\wedge \mathrm{vol}_{\mathbb{T}^3} \ ,
\\[1mm]
B_{(6)} =&\ 4\ell^2 H\left({Q_5}^{-1} K^2 r \, \mathrm{vol}_{\mathrm{Mkw}_2}\wedge dx_2\wedge\mathrm{vol}_{\mathbb{T}^3}-2\ell^2{Q_5} K^{-1}\mathrm{vol}_{\mathrm{AdS}_3}\wedge dx_1 \wedge \mathrm{vol}_{S^2}\right)\ ,
\\[1mm]
e^{\Phi}=&\ H^{-1}\ , \label{AdS32_dilaton}
\end{align}
where $ds^2_{\mathrm{AdS}_3}\equiv \frac{Q_5}{\ell}\frac{r}{\ell}ds^2_{\mathrm{Mkw}_2}+\frac{dr^2}{4r^2}$ is the metric of unit AdS$_3$, with $\mathrm{vol}_{\mathrm{AdS}_3}\equiv \frac{Q_5}{2\ell^2}\, dt\wedge dy\wedge dr$ the volume element, $ds_{\Sigma}^2\equiv\left(dx_1^2+dx_2^2\right)$ and  $ ds_{\mathbb{T}^3}^2\equiv\left(d\rho^2+\rho^2  ds_{\tilde{S}^2}^2\right)$, while the constant length $\ell$ is defined as $\ell\equiv (Q_{\mathrm{D}3}Q_{5}^3)^{1/4}$, and finally $H$ \& $K$ are two arbitrary harmonic functions on $\Sigma$ and $\mathbb{T}^3$, respectively.

The above solution preserves $8$ real supercharges and represents $\mathrm{AdS}_3\times S^2$ warped over $\mathbb{T}^3\times\Sigma$. In particular, for a spherically symmetric choice for the function $K$, the resulting geometry can be thought of as $\mathrm{AdS}_3\times S^2\times \tilde{S}^2$ warped over $I_{\rho}\times\Sigma$.
\subsection*{Discussion}
The solution \eqref{AdS32_metric}  interpreted as $\mathrm{AdS}_3\times S^2\times \tilde{S}^2$ warped over $I_{\rho}\times\Sigma$ is a straightforward generalization of the one discussed in \cite{Faedo:2020lyw}. 
\footnote{Let us note that, in contrast with the brane intersection of \cite{Faedo:2020lyw}, in this solution the D7 brane can be lobectomized taking $H_{\mathrm{D}7}\to 1$.}
The generalization consists in taking $H$ as a general harmonic function on $\Sigma$, rather than just being linear in one of the two coordinates, say $x_1$. As a consequence, since we have no Abelian isometries left in our background, we conclude that our setup is \emph{genuinely} type IIB, with no relation to (massive) type IIA any longer. On the other hand though, the physical interpretation proposed there is directly carried through in our more general situation. Let us summarize in what follows the salient steps.

In particular, one may use the existence of a consistent truncation of type IIB supergravity on $S^2\times \Sigma$ \cite{Hong:2018amk} to land within minimal $\mathcal{N}=(1,1)$ $D=6$ gauged supergravity \cite{Romans:1985tw}. This theory possesses 16  real supercharges and $\mathrm{SU}(2)$ gauge symmetry, and is often referred to as \emph{Romans' supergravity}.  Its field content is given by the 6d metric, a real scalar field $X$, a 2-form gauge potential $\mathcal{B}_{(2)}$, three $\mathrm{SU}(2)$ vector fields plus one Abelian vector field.  The scalar potential of the theory admits a real superpotential formulation given as
\begin{equation}
V(X) \ = \ 16\,\left(-5f(X)^2+X^2\left(D_Xf\right)^2\right) \ ,
\end{equation}
where $f(X) \, \equiv \, \frac18\left(3X+X^{-3}\right) $. Note that the theory admits a SUSY AdS$_6$ extremum at $X=1$ in the absence of vectors and 2-form.

Now, by using the uplift formulae in \cite{Hong:2018amk}, one can show that, upon specifying a choice of harmonic function on $\Sigma$, the SUSY vacuum of the minimal 6D theory lifts to the class of AdS$_6$ vacua described in \cite{DHoker:2016ujz,DHoker:2017mds,DHoker:2017zwj}, where a brane interpretation is provided in terms of $(p,q)5$ brane webs and D$7$ branes. Furthermore, 6D BPS slicings of the form $\mathrm{AdS}_3\times S^2$ were studied in  \cite{Dibitetto:2018iar}. The authors of \cite{Faedo:2020lyw} were able to map the solutions obtained when lifting these configurations to type IIB to the near-horizon limit of D3 -- D5 -- NS5 -- D5$'$ -- NS5$'$ -- D7 brane systems for which $H_{\mathrm{D}7}$ is purely linear in $x_1$.

This implies that our solution \eqref{AdS32_metric} may be interpreted holographically as the gravity dual of a defect CFT$_2$ inside a 5d $\mathcal{N}=1$ SCFT, which is engineered by placing within the background $(p,q)5$ web and D7 branes, D3 branes and a new $(p,q)5$ web only sharing two directions with the previous one. The more general $(x_1,x_2)$ dependence present in our solution w.r.t. the one in \cite{Faedo:2020lyw} simply suggests that the AdS$_6$ vacuum, which is reached asymptotically when moving at infinite distance from the defect branes, will no longer necessarily correspond to the \emph{annulus} as a choice for $\Sigma$.

\section*{Acknowledgments}

We would like to thank Yolanda Lozano, Nicol\`o Petri and Shigeki Sugimoto for some valuable comments on a draft version of this manuscript.
The work of JRB is supported by Fundación S\'eneca, Agencia de Ciencia y Tecnolog\'ia de la Regi\'on de Murcia, grant 21472/FPI/20.
The work of GD is supported by the STARS grant named THEsPIAN.
 The work of JJFM is supported by Universidad de Murcia-Plan Propio Postdoctoral, the Spanish Ministerio de Econom\'ia y Competitividad and CARM Fundaci\'on S\'eneca under grants FIS2015-28521 and 21257/PI/19.

\appendix
\section{Type IIB supergravity in the string frame}
\label{app:iib-string}

Type IIB supergravity in its democratic formulation \cite{Bergshoeff:2001pv} is described in terms of the common NS-NS sector $\{g, B_{(2)},\Phi\}$ coupled to even form fields $\{C_{(2p)}\}_{p=0,1,2,3,4}$. The (bosonic) dynamics of the theory can be derived from the following \emph{pseudoaction}\footnote{Throughout this paper we will retain conventions such that $\kappa_{10}=1.$}
\begin{equation}
\label{IIB_action}
S_{\mathrm{IIB}} \, = \, \frac{1}{2\kappa_{10}^2}\int{d^{10}x\sqrt{-g}\left(e^{-2\Phi}\left(\mathcal{R}+4(\partial\Phi)^2-\frac{1}{12}|H_{(3)}|^2\right)-\frac{1}{4}\sum\limits_{p=0}^{4}\frac{|G_{(2p+1)}|^2}{(2p+1)!}\right)} \ ,
\end{equation}
where the appearing (modified) field strengths read
\begin{align}
H_{(3)} =&\ dB_{(2)}\ , \\
H_{(7)}=&\ dB_{(6)}+\frac{1}{2}(-G_{(7)}\wedge C_{(0)}+G_{(5)}\wedge C_{(2)}-G_{(3)}\wedge C_{(4)}+G_{(1)}\wedge C_{(6)})\ , \\
G_{(1)}=&\ dC_{(0)} \ , \\ 
G_{(3)} =&\  dC_{(2)}-H_{(3)}\wedge C_{(0)} \ , \\
G_{(5)}=&\ dC_{(4)}-H_{(3)}\wedge C_{(2)}\ ,\\
G_{(7)}=&\ dC_{(6)}-H_{(3)}\wedge C_{(4)}\ ,\\
G_{(9)}=&\ dC_{(8)}-H_{(3)}\wedge C_{(6)}\ ,
\end{align}
which are designed to automatically satisfy the following (modified) Bianchi identities (BI)
\begin{align}
dH_{(3)}=&\ 0\ ,\label{BIH3} \\
dH_{(7)}+\frac{1}{2} \sum\limits_p \star G_{(p)} \wedge G_{(p-2)} =&\ 0\ , \\[-2mm]
dG_{(1)}=&\ 0\ ,\\ 
dG_{(3)}-H_{(3)}\wedge G_{(1)}=&\ 0\ ,\\
dG_{(5)}-H_{(3)}\wedge G_{(3)}=&\ 0\ ,\\ 
dG_{(7)}-H_{(3)}\wedge G_{(5)}=&\ 0\ ,\\
dG_{(9)}-H_{(3)}\wedge G_{(7)}=&\ 0\ .\label{BIG9}
\end{align}
It is worth mentioning that $S_{\mathrm{IIB}}$ in \eqref{IIB_action} is called a pseudoaction because it must be supplemented by the following \emph{duality relations}
\begin{align}
G_{(9)}\overset{!}{=}\star G_{(1)}
\ ,
\quad
G_{(7)}\overset{!}{=}-\star G_{(3)}
\ ,
\quad
G_{(5)}\overset{!}{=}\star G_{(5)}
\ ,
\quad
\ H_{(7)}\overset{!}{=}e^{-2\Phi}\star H_{(3)}
\ ,
\end{align}
that yield the correct number of propagating degrees of freedom and hence allow for an on-shell realization of supersymmetry.
By varying  \eqref{IIB_action}, one obtains the following set of equations of motion
\begin{equation}\label{EOMPhi}
\Box\Phi \,-\, (\partial\Phi)^2\,+\,\frac{1}{4}\,\mathcal{R}\,-\,\frac{1}{8\times 3!}\,|H_{(3)}|^2\, = \, 0 \ ,
\end{equation}
for the 10D dilaton $\Phi$,
\begin{align}
& d(e^{-2\Phi}\star H_{(3)})+\frac{1}{2}\sum\limits_p \star G_{(p)}\wedge G_{(p-2)}=\ 0\ , & d(e^{2\Phi}\star H_{(7)})=  \ 0\ , \\
& d(\star G_{(1)})+H_{(3)}\wedge(\star G_{(3)})= \ 0\ , & d(\star G_{(3)})+H_{(3)}\wedge G_{(5)}=\ 0\ ,\\[3mm]
& d(\star G_{(5)})=d G_{(5)}= H_{(3)}\wedge G_{(3)} \ ,
\end{align}
for the form fields, and finally the (trace reversed) Einstein equations
\begin{align}\label{EinsteinEQ}
0=& \ e^{-2\Phi}\left(
	\mathcal{R}_{MN}
	+2\nabla_M\nabla_N\Phi
	-\frac{1}{4} H_{MPQ} H_N{}^{PQ}
	\right)
-\frac{1}{2} (G_{(1)}^2)_{MN} \notag\\
&-\frac{1}{2\times 2!} (G_{(3)}^2)_{MN}
-\frac{1}{4\times 4!} (G_{(5)}^2)_{MN}
+\frac{1}{4}g_{MN}\left(
	|G_{(1)}|^2
	+\frac{1}{3!}|G_{(3)}|^2
	\right)
\ .
\end{align}

\section{Intersecting type IIB BPS objects and SUSY projectors}
\label{app:projectors}

In $(1+9)$D, spinors admit a Majorana-Weyl (MW) representation in which the Dirac matrices are realized as
\begin{align}
\Gamma^M \ = \ \left(\begin{array}{c|c} 0_{16} & \Sigma^M \\  \overline{\Sigma}^M & 0_{16}\end{array}\right) \ ,
\end{align}
where the $16\times 16$ blocks $ \Sigma^M$ \& $\overline{\Sigma}^M$ are real and act on chiral spinors. Since the 32 real supercharges of type IIB supergravity are represented by two distinct MW spinors of the same chirality, the same degrees of freedom can be rearranged into a single \emph{complexified} chiral spinor
\begin{equation}
\epsilon \ = \ \zeta \,+\, i \, \eta \ ,
\end{equation}
where both $\zeta$ and $\eta$ are MW. In terms of such complex MW spinor, $\frac{1}{2}$-BPS objects admit sets of Killing spinors spanned by the eigenspinors of a projection operator 
\begin{equation}
\label{SUSY_proj}
\Pi(\mathcal{O}) \ \equiv \ \frac{1}{2}(\mathbb{I}\,+\,\mathcal{O}) \ ,
\end{equation}
where the operator $\mathcal{O}$ is an involution acting on complex spinors such as $\epsilon$. The different SUSY projectors for the various fundamental BPS objects in type IIB supergravity are summarized in table \ref{table:SUSY_projectors}.
\begin{table}[http!]
\renewcommand{\arraystretch}{1}
\begin{center}
\scalebox{1}[1]{
\begin{tabular}{|c||c| c| c| c | c| c| c| c| c|}
\hline
object & F$1$ & NS$5$ & W$_{\textrm{B}}$ & KK$5_{\textrm{B}}$ & D$1$ & D$3$ & D$5$ & D$7$ & D$9$ \\
\hline
$\mathcal{O}$ & $\Sigma^{01}\circ *$ & $\Sigma^{6789}\circ *$ & $\Sigma^{01}$ & $\Sigma^{6789}$ & $i\Sigma^{01}\circ *$ & $i\Sigma^{0123}$ & $i\Sigma^{6789}\circ *$ & $i\Sigma^{89}$ & $i *$\\
\hline
\end{tabular}
}
\end{center}
\caption{The different operators $\mathcal{O}$ appearing in the SUSY projector defined in \protect\eqref{SUSY_proj} for the various fundamental BPS objects of type IIB string theory. We introduced the notation  $\Sigma^{i_1...i_{2p}}\,\equiv\,\Sigma^{[i_1}\cdots\overline{\Sigma}^{i_{2p}]}$, while $*$ denotes complex conjugation. All objects are assumed to fill the first $(p+1)$ spacetime directions.} \label{table:SUSY_projectors}
\end{table}

When studying intersections of fundamental BPS objects, one has to find a set of common eigenspinors simultaneously preserved by all projectors corresponding to the various objects involved. Note that this will only be possible if the aforementioned operators commute. The (real) dimension of the common eigenspace will then represent the number of preserved supercharges.

Let us briefly discuss the relevant cases for this paper.
\begin{itemize}
\item \textbf{D3 -- D5 -- D7:} This is the brane intersection illustrated in table \ref{Table:D3D5D7}. It may be seen to preserve four real supercharges by simultaneously studying the Killing spinor projections associated with each of the branes appearing. This translates into
\begin{align}
\Pi(\mathcal{O}_{\mathrm{D}3})\, \epsilon \ \overset{!}{=} \ \Pi(\mathcal{O}_{\mathrm{D}5})\, \epsilon \ \overset{!}{=} \ \Pi(\mathcal{O}_{\mathrm{D}7})\, \epsilon \ \overset{!}{=} \ \epsilon \ ,
\end{align}
which admits a space of solutions parametrized by four independent real parameters, thus giving rise to a $\frac{1}{8}$ BPS brane intersection.

\item \textbf{D3 -- D5 --NS5 -- D5$'$ -- NS5$'$ -- D7:} Starting from the previous configuration, one might add D5 and NS5 branes as shown in table \ref{Table:D3D5NS5D5NS5D7}. Interestingly, even though one might expect this configuration to be way less supersymmetric than the previous one, in the end none of the new projections are really independent of the ones already taken above. As a result, a generic Killing spinor found earlier will also be a good spinor for the new intersection of branes, which is therefore \emph{still} $\frac{1}{8}$ BPS.

\item \textbf{D3 -- D3$'$ -- D7:} This situation is the one considered in section \ref{Sec3}. By imposing
\begin{align}
\Pi(\mathcal{O}_{\mathrm{D}3})\, \epsilon \ \overset{!}{=} \ \Pi(\mathcal{O}_{\mathrm{D}7})\, \epsilon \ \overset{!}{=} \ \Pi(\mathcal{O}_{\mathrm{D}3'})\, \epsilon \ \overset{!}{=} \ \epsilon \ ,
\end{align}
we realize that the three projections are perfectly compatible and each of them halves the space of invariant eigenspinors in such a way that the resulting solution again depends on four real parameters. This means that this brane system is $\frac{1}{8}$ BPS, as well.
\end{itemize}

 \bibliographystyle{utphys}
  \bibliography{AdS3_IIB}
\end{document}